# Multi-level Resistive Switching Characteristics of W/Co:TiO$_2$/FTO Structures

Haitao Tang, Zhi Luo, Zhao Yang, Bin Yang, Bo Huang, and Weiguang Xie

*Abstract*—**In the present work, multi-level resistive switching (RS) in W/Co:TiO$_2$/FTO structures induced by a multi-mixed mechanism was studied. It was found that the devices could be reproducibly programmed into three nonvolatile resistance states. And the directly switching between any resistance states was realized. This increases the operation speed and lowers the complexity of control circuit of multi-state nonvolatile memory.**

## I. INTRODUCTION

Oxide-based resistance random access memory (RRAM), which is very promising as next generation nonvolatile memory (NVM), has attracted considerable interest recently. Previous studies showed that conductive bridge, valence change and charge trapping place important role in resistive switching (RS). The RS characteristics of devices is determined by one or more mechanisms. However, for a given device, it is difficult to control the determining mechanism. In order to solve this problem, it is necessary to clarify the underlying physics of RS.

Multi-level RS, which induced either by limiting the current compliance (CC) or by controlling the reset voltage, has been reported by several groups. In reset voltage control mode, various resistance states come from the variation of filament length. On the other hand, in CC control mode, various resistance states originate from the variation of filament width. However, in most of these Multi-level RS devices, to switch from one intermediate resistance state (IRS1) to another intermediate resistance state (IRS2), the device has to go back to HRS (or LRS) from IRS1 and then jump to IRS2. These processes decrease the operation speed and increase the complexity of control circuit.

In this paper, W/Co:TiO$_2$/FTO devices were fabricated and their RS characteristics were investigated. Very good multi-level RS behavior was observed, and three resistance states could be transformed into each other freely. Detail analysis indicated that the multi-level RS originates from a multi-mixed effect of W/Co:TiO$_2$ Schottky-like interface and the formation/rupture of oxygen vacancy filaments in Co:TiO$_2$ film.

## II. EXPERIMENTS

Co:TiO$_2$ thin films were fabricated by sol-gel spin coating technique. TiO$_2$ sol was prepared by a modified hydrolysis process. Co(CH$_3$COO)$_2$ 4H$_2$O was used as the doping source and mixed with the sol at a molar ratio of Co:Ti=1:100, which corresponds to a nominal doping concentration of about 1 at. %. Thin films with desired thicknesses (about 120 nm) were deposited onto FTO glass followed by annealing in air ambient at 550 ℃ for 30 min. A rounded tungsten tip was used as top electrode.

All devices were electrically characterized in the dark with computer controlled Keithley 2400 at room temperature. A forward (positive) bias applied to the device was defined as the current flowing from the top W electrode into the thin film.

## III. RESULTS AND DISCUSSION

The multi-level RS properties of TiO$_2$ thin film with W/Co:TiO$_2$/FTO configuration were investigated. As shown in Fig. 1, the directly switching between any resistance states of the device was realized. 1) Switching between HRS and LRS (Fig. 1(a)). The sequence of voltage-sweeping were shown as arrows 1 to 4 in Fig. 1(a). A switching from HRS to LRS was achieved by applying a negative voltage of -3.4 V ($V_{set1}$) with a CC of 4 mA. Subsequently, the reset process which switched the device from LRS to HRS was performed by sweeping the negative voltage to −1 V ($V_{reset1}$) without CC. 2) Switching between HRS and intermediate resistance state (IRS) (Fig. 1(b)). The positive set process which switched the device from HRS to IRS was observed by applying a positive voltage of 2.6 V ($V_{set2}$) with a CC of 5 mA. One obvious difference between IRS and LRS is that the I-V curves is linear in LRS while it is nonlinear in IRS. This implies that the transport mechanism is different in these two states. And the IRS could be reset to HRS by applying a negative bias ($V_{reset2}$) without CC. 3) Switching between IRS and LRS (Fig. 1 (c)). A bipolar switching between IRS and LRS could be achieved with a CC of 5 mA. If the device was set to IRS with a positive bias ($V_{Set2}$), it would transfer to LRS under a negative bias of about -1 V ($V_{set3}$), and

This paragraph of the first footnote will contain the date on which you submitted your paper for review. It will also contain support information, including sponsor and financial support acknowledgment. This work has been supported by the the National Natural Science Foundation of China under Grant 11204105. W.G. Xie would like thanks the financial support of Guangdong Natural Science Foundation (Grants Nos. 2014A030313381).

H. T. Tang, Z. Luo, Z. Yang, B. Yang and B. Huang are with the Department of Electronic Engineering, Jinan University, Guangzhou, Guangdong, 510632, PR China (e-mail: zhluocn@gmail.com)

W.G. Xie is with Department of Physics, Siyuan Laboratory, Jinan University, Guangzhou, Guangdong, 510632, PR China.

switch back to IRS under a positive bias of about 0.7 V ($V_{reset3}$).

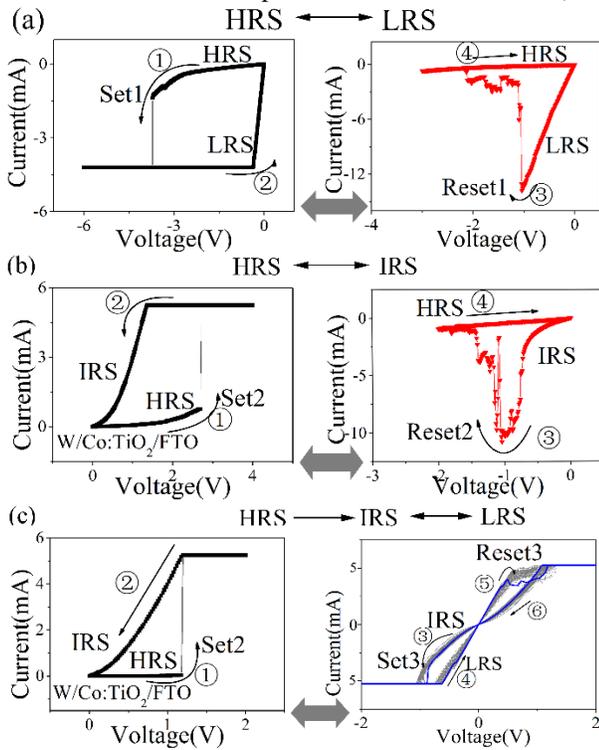

Fig. 1. (a) Switching between HRS and LRS. (b) Switching between HRS and IRS. (c) Switching between IRS and LRS.

Although the switching voltage and resistance in different samples are different and these parameters may change slightly in different cycle even in the same device, most devices showed similar RS characteristics. It should also be emphasized that all of these switching can be triggered by different control strategy independently. For example, a device in HRS can be switched to LRS by applying a negative bias with a CC of 5 mA, or to IRS by applying a positive bias with a CC of 5 mA. Fig. 2 (a) showed the endurance times of the devices, and the cycles of resistance states were programmed to HRS→ IRS→ LRS→ HRS…. This clearly demonstrated that the device switching of LRS, IRS, and HRS could reach up to 50 times. Fig. 2 (b) showed that all the three resistance states could retain for more than 1000 seconds. We believed that the resistance value of different states could be further stabilized and the endurance performance could be improved by optimizing the device fabrication parameters.

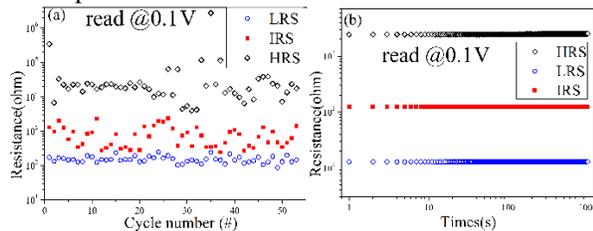

Fig. 2. (a) Endurance properties of W/Co:TiO$_2$/FTO structure with a reading voltage of 0.1 V. (b) Retention of the HRS and LRS at RT in the W/Co:TiO$_2$/FTO structure with a reading voltage of 0.1 V.

The I-V curves in Fig. 1 were analyzed to clarify the mechanism in the device. The I-V curves in HRS was fitted with the trap-filled space charge limited current (SCLC) (Fig. 3 (a)). In the low voltage region ranging from 0 to -1 V, the I–V relationship showed a linear dependence on voltage, suggesting an Ohmic conduction region. In middle voltage region from -1.4 to 2.6 V, the current obeyed the square dependence on voltage, which is accorded with the Child's square law ($I \sim V^2$). While at the higher voltage region, the current increased faster, correspondingly to the steep increase region in trap-controlled SCLC. The fitting results of I−V curves in IRS of the voltage region ranging from 0 to -0.7 V were shown in Fig. 3(b). The I−V curves obeyed the linear relation of ln(I) versus $(V)^{1/2}$, indicating that the conduction mechanism was dominated by Schottky emission. And it was clear that the I-V curves in LRS followed Ohm's law (Fig. 3(c)). As the transport mechanism of HRS, IRS and LRS is different, the switching mechanisms between these states may be different too.

The multi-level resistive switching in W/Co:TiO$_2$/FTO may originate from a multi-mixed mechanisms, as schematically depicted in Fig. 4. At the beginning, the initial device was in HRS. Oxygen vacancy distributed in Co:TiO$_2$ randomly (Fig. 4(a)). In general, Co:TiO$_2$ is an n-type semiconductor, so the depletion layer is with positively charge. 1) When a positive voltage ($V_{Set2}$) was applied on the top electrode (TE) with a CC of 5 mA, Oxygen vacancy metallic filaments were formed via oxygen vacancies drift and deposition, and the depletion layer became narrower with the reduced Schottky-like barrier at the W/Co:TiO$_2$ interface. But the filaments could not reach the W anode because of the repulsive force between the positively charged oxygen vacancy and the depletion layer. Therefore, the device was switched to IRS(Fig. 4(b)). Then the applied bias polarity was reversed to $V_{Set3}$ with a CC of 5 mA, the oxygen vacancies which were driven by electric field were deposited on TE against the charge repulsion, even if the barrier was increased and the depletion layer was broadened. Oxygen vacancy metallic filaments were generated from TE to BE (Fig.4(c)). 'Set3' process occurred, and the LRS was achieved. The LRS follows the Ohmic behavior, and is consistent with the typical filamentary model. If the bias polarity reversed to $V_{Reset2}$ without CC after 'Set2' process，oxygen vacancy metallic filaments would be generated from TE to BE at first, and then the filaments were ruptured due to the heat which was generated by the large current flow (Fig.4(d)). This was 'Reset2' process, and the process switched the device from IRS to HRS. Repeated the 'Set2' process, oxygen vacancies in the interface were extruded. The device was switched from HRS to IRS. 2) When a negative voltage ($V_{Set1}$) was applied on the TE of the initial device with a CC of 5 mA, the oxygen vacancies tended to drift toward TE. Oxygen vacancy metallic filaments were formed directly from TE to BE. Therefore, the device was switched from HRS to LRS ('Set1'). Then the bias polarity was reversed to $V_{Reset3}$ with a CC of 5 mA, oxygen vacancies in the interface were extruded. As a result the device was switched from LRS to IRS ('Reset3'). If the negative bias voltage ($V_{Reset1}$) without CC was applied on the TE after 'Set1' process, oxygen vacancy metallic filaments would be ruptured by the heat ('Reset1'), the device was switched from LRS to HRS. Repeated the 'Set1' process, Oxygen vacancy metallic filaments were generated from TE to BE again, the device was switched from HRS to LRS.

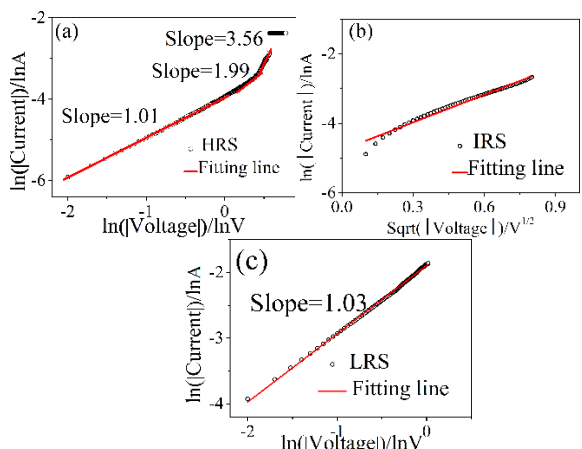

Fig. 3. The fitting results of I-V curve in HRS (a), IRS (b) and LRS (c).

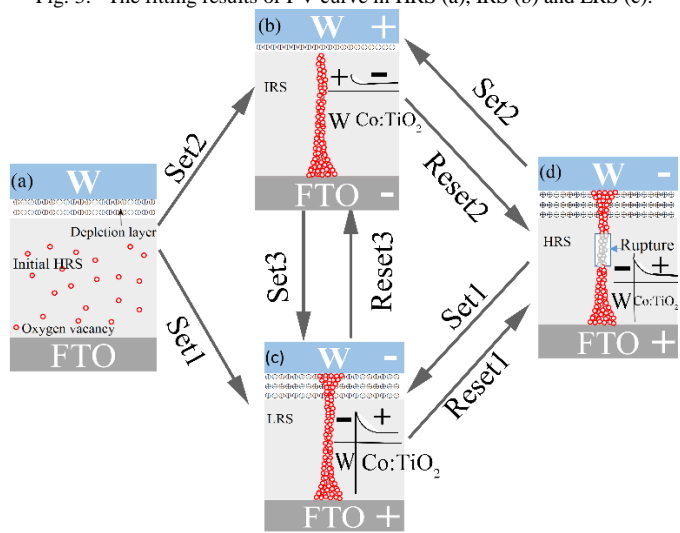

Fig. 4. A schematic illustration for the multi-mixed mechanism of the multi-level resistive switching in W/Co:TiO$_2$/FTO structures. (a) Initial state. The cell exhibiting randomly distributed oxygen vacancies. (b) IRS. oxygen vacancies drift toward BE and deposit. (c) LRS. Oxygen vacancy metallic filaments formed. (d) HRS. Oxygen vacancy filaments dissolved.

## IV. CONCLUSION

In summary, multi-level RS was observed in W/Co:TiO$_2$/FTO structures. The devices showed good multi-level RS performance, and three resistance states could be transformed into each other easily. Detailed analysis of I-V curves in HRS, IRS and LRS suggested that the multi-level RS events were induced by a multi-mixed effect of Schottky-like interface between W and TiO$_2$ films and the formation/rupture of oxygen vacancy filaments.